\documentclass{article}

\usepackage{times}
\usepackage{graphicx} 
\usepackage{subfigure} 

\usepackage{natbib}

\usepackage{algorithm}
\usepackage{algorithmic}

\usepackage{hyperref}

\usepackage[accepted]{icml2017}

\usepackage{graphicx}
\graphicspath{ {images/} }

\icmltitlerunning{Gradient Augmented Information Retrieval with Autoencoders and Semantic Hashing}

\begin{document} 

\twocolumn[
\icmltitle{Gradient Augmented Information Retrieval with Autoencoders and Semantic Hashing (L101 Michaelmas 2017)}




\begin{icmlauthorlist}
\icmlauthor{Sean Billings}{cam}
\end{icmlauthorlist}

\icmlaffiliation{cam}{University of Cambridge, Cambridge, United Kingdom}

\icmlcorrespondingauthor{Sean Billings}{sb2219@cam.ac.uk}

\icmlkeywords{boring formatting information, machine learning, ICML}

\vskip 0.3in
]



\printAffiliationsAndNotice{}  

\begin{abstract} 

This paper will explore the use of autoencoders for semantic hashing in the context of Information Retrieval. This paper will summarize  how to efficiently train an autoencoder in order to create meaningful and low-dimensional encodings of data. This paper will demonstrate how computing and storing the closest encodings to an input query can help speed up search time and improve the quality of our search results. The novel contributions of this paper involve using the representation of the data learned by an auto-encoder in order to augment our search query in various ways. I present and evaluate the new gradient search augmentation (GSA) approach, as well as the more well-known pseudo-relevance-feedback  (PRF) adjustment. I find that GSA helps to improve the performance of the TF-IDF based information retrieval system, and PRF combined with GSA works best overall for the systems compared in this paper. \\

\end{abstract}

\section{Introduction}

Nearly every internet user is familiar with web search. The field of research underlying these kinds of systems is known as Information Retrieval (IR). For this paper, I explore the task of finding the most similar documents in a collection to a given search query. There are some fairly standard IR techniques such as TF-IDF that can be used to tackle these problems, but the standard approaches often fail in terms of accuracy and in terms of scalability on large enough datasets. \\

In \cite{semantichashing}, a new approach to intelligently encode and store a collection of documents using autoencoders is proposed. When combined with standard techniques such as TF-IDF,  the authors of that paper find both speed and performance improvements. This paper hopes to extend upon previous work and better explore both the theory underlying why autoencoders are useful in information retrieval tasks and also to explore new techniques to help further improve performance. The theory regarding the training of autoencoders is somewhat extensive, so this paper includes a sufficient review of why autoencoders are useful, and how to train them. This paper will also explain the standard TF-IDF methodologies, and known improvement strategies such as pseuodo-relevance-feedback.\\

Section 2 of this paper will cover the theory underlying autoencoders. This will include discussion of training methodologies using RBM as proposed in \cite{hintonrbmpractical}, and also more typical back-propagation approaches \cite{bishopnn} that have been favoured in the contemporary neural network toolboxes such as Keras. \\

Section 3 will cover the core methodology of semantic hashing. We will discuss the benefits proposed in \cite{semantichashing}. The main engineering parameters of this system will be discussed including exploration of how tuning the various semantic hashing system components can affect performance. \\

Section 4 will describe several Information Retrieval techniques, which are to be evaluated. This discussion will include the known techniques such as TF-IDF and pseudo relevance feedback techniques, and also the new gradient search augmentation approach presented in this paper. \\

Section 5 summarizes and presents experiments that explore the results of the novel techniques of gradient search augmentation and the existing pseudo-relevance-feedback. These will be compared to the semantic hashing TF-IDF variant baseline. \\

Section 6 concludes this paper with a discussion of the merits of semantic hashing and a summary of the relevant findings in this paper. \\

\section{Training Autoencoders} 
 
In this section we develop the theory that underlies autoencoders and show how these types of neural networks can be used to encode a useful structure through which to represent a set of documents. \\

\subsection{Autoencoder theory}

Autoencoders are a form of neural network that train a reconstruction function $ r=g(f(x)) $ using some loss function $ L(x,r(x)) $ in order to produce a more informative and/or lower dimensional reconstruction of our data \cite{deeplearningbook}. Whereas neural networks are most commonly used to generate a high-dimensional parameterized function in order to solve a discriminative task, autoencoders are trained to regenerate the data that the model is trained on. If we perform regularization to our loss function and do strategic structural adjustment to our network, the autoencoder learns a manifold structure $ \frac{\partial \log p(x)}{\partial x}$ for the probabilistic density $ p(x) $ of the training data \cite{bengiotheory}. \\

More formally, let us consider a neural network that is trained as an autoencoder. As in \cite{bengiotheory}, let us consider a Denoising Auto Encoder (DAE) that aims to generate a simplified reconstruction r(x) by minimising the criterion $L_{DAE} = E[ L(x,r(N(x))) ] $ where $N(x) = x + \epsilon$ for some gaussian error $ \epsilon = N(0,1)$. It can be shown that we produce the following optimal reconstruction function $r^*(x) =  x + \sigma^2 \frac{\partial \log p(x)}{\partial x} + o(\sigma^2) $ which behaves asymptotically with the variance of the gaussian noise as $\sigma \rightarrow 0$. The asymptotic behavior $ \frac{r_\sigma(x)-x}{\sigma^2} \rightarrow \frac{\partial \log p(x)}{\partial x}$ as $\sigma \rightarrow 0$ allows us to determine the informative structure of x via the induced manifold on p(x). \\

\subsection{Training deep autoencoders with back propagation}

Neural networks can be broken down into the combination of an input layer, a stack of hidden layers, and an output layer. The input layer connects an input vector x to our first hidden layer along with a bias term. Each hidden layer at depth k of our network is parameterized by a set of weights $ w^k_{ij} $ defining a set of activations $a^k_j = \sum w_{ij}z^k_i $ and a set of nonlinear functions $g^h(a_{j}) $ that act on those activations in order to generate the input $z^{k+1}_i$ to the next layer of our network. For regression problems, the output layer of our neural network computes a single real value $ \hat{y} $ that minimizes a loss function $ L(y(x),\hat{y}) $. For an autoencoder, the output layer of our neural network generates the reconstructed vector $ r(x) $, which minimizes a loss function $ L(x,r(x)) $. \\

\begin{algorithm}[tb]
   \caption{Backpropagation}
   \label{alg:example} 
\begin{algorithmic}
   \STATE {\bfseries Input:} Network \{network structure g , initial weights w\}, learning rate $\eta$, $\{x_n\}$ training examples
   \STATE {\bfseries Output:} Network \{weights $w^*$\}
   \FOR{ $iterations$}
   \FOR{ $x_n$ in training examples}
   \STATE Forward pass to compute $ E_n $
   \STATE Backward pass to compute $\nabla E_n = \frac{\partial E}{\partial w^k_{ij}} $
   \STATE update $ w^k_{ij} = w^k_{ij} - \eta \nabla E_n$
   \ENDFOR
   \ENDFOR
\end{algorithmic}
\end{algorithm}

If we explicitly define our loss function we can then compute an error $E_n = L(x,r(x_n))$ for each data point $x_n$ in our data set. The backpropagation algorithm (see Algorithm 1) works by performing a forward pass over the network to compute the errors $ E_n $, and then a backwards pass over the network in order to compute the derivatives $ \frac{\partial E_n}{\partial w^k_{ij}}$ for each layer of the network. \\

Formally, we can break down the training of a denoising autoencoder (DAE) with backpropagation as follows. We define the error function for our denoising autoencoder as the euclidian distance for the input against the reconstruction of the input with noise 

\begin{equation}
E_{DAE} =  \frac{1}{2} (x - r(N(x))^2)
\label{eq:errordae}
\end{equation}

Using the chain rule, we can derive the form of the derivative of the error function with respect to the weights as 

\begin{equation}
\frac{\partial E_{DAE}}{\partial w^k_{ij}}=  \frac{\partial E_{DAE}}{\partial a^k_{j}} \frac{\partial a^k_{j}}{\partial w^k_{ij}} 
\label{eq:ederivativedae}
\end{equation}

Where $ \frac{\partial a^k_{j}}{\partial w^k_{ij}} = z^k_i$. As in \cite{bishopnn}, we use the notation $\delta^k_j = \frac{\partial E_{DAE}}{\partial a^k_{j}} $. This is used for convenience to represent the error differences at layer $k$. This allows us to explicitly represent and compute the error difference computations at each layer of our neural network. 

For the output layer, we compute the error differences as

\begin{equation}
\delta^k_j = g^{k'}(a_j) \frac{1}{2} \frac{\partial (x - r(N(x))^2)}{\partial r(N(x))}
\label{eq:outputdj}
\end{equation}

For the hidden layers, we compute the error differences as

\begin{equation}
\delta^k_j = g^{k'}(a_j) \sum_i w^{k+1}_{ij} \delta^{k+1}_i
\label{eq:outputdj}
\end{equation}

where we iterate over the nodes $i$ in the $k+1$ layer that the node $j$ in the kth layer is connected to.\\

As shown above, the gradient of one layer depends solely and simply on the activation functions and the error of the next layer. These derivatives are fed into a gradient based optimization algorithm in order to train the neural network. These gradients can be computed efficiently, especially when taking into consideration GPU hardware. However, the number of iterations required to train a strong network can still be very high, and a lot of compute power can be required. For this reason, in the following section we consider an alternative method for training neural networks that has been shown to work sufficiently well for training autoencoders. \\

\subsection{Training autoencoders with restricted boltzmann machines}
 
Training an autoencoder as a stack of resricted boltzmann machines (RBM) can be seen as a greedy approach to training a neural network \cite{hintonrbmpractical}. In this section, I will describe how a stack of RBM can be trained to approximate the structure learned via backpropagation and how they can be used as an alternative. This is the approach used in \cite{semantichashing}. 

Let us first develop the theory underlying a single restricted boltzmann machine. A single RBM consists of a single layer of probabilistic hidden units h, connected to a single layer of usually binary visible units with weighted connections. The RBM learns a joint probability distribution $ p(v,h)$. where 
 
 \begin{equation}
p(v,h) = \frac{1}{Z} e^{-E(v,h)}
\label{eq:daeloss}
\end{equation}
 
 The distribution over the visible and hidden units is determined by an energy function $E$ and the partition function $Z$. An example energy function for an RBM with visible unit biases $a_i$ hidden unit biases $ b_j$, and linear hidden to visible unit activations is 
 
 \begin{equation}
- E(v,h) = \sum_{visible} a_i v_i + \sum_{hidden} b_j h_j + \sum _{i,j} v_i h_j w_{ij}
\label{eq:energyrbm}
\end{equation}

The partition function $Z$ is then defined as the sum over the energy configurations of the potential settings of the visible and hidden units.

 \begin{equation}
Z = \sum_{v,h} e^{-E(v,h)}
\label{eq:pnodevector}
\end{equation}

The weights of the RBM can be learned via the Constrastive Divergence (CDn) algorithm (see Algorithm 2). The CDn algorithm is often parametized by n, where n is the number of samples taken at each step of the algorithm. There are also more elaborate and efficient variations of CDn, such as the less intuitive Persistent Constrastive Divergence (PCD) algorithm. PCD is the best and most efficient algorithm to use in practice due to its superior learning properties. 

\begin{algorithm}[tb]
   \caption{Contrastive Divergence (CD1)}
   \label{alg:example} 
\begin{algorithmic}
   \STATE {\bfseries Input:} RBM \{v, h, w\}, learning rate $\eta$
   \STATE {\bfseries Output:} RBM  \{v, h, $w^*$ \}
   \FOR{ $k=1$ {\bfseries to} $iterations$}
   \STATE compute $p(h_j=1| v) =  \sigma (b_j + \sum_i v_i w_{ij} )$ for all hidden units $h_j$
   \STATE set $h^*_j = sample(p(h_j | v))$
   \STATE compute $<v_i h*_j>_{data}$
   \STATE compute $p(v_i=1| h^*) =  \sigma (a_i + \sum_j h_j w_{ij} )$ 
   \STATE set $v^*_i = sample(p(v_i | h^*))$
   \STATE compute $<v*_i h^*_j>_{model}$
   \STATE set $ \Delta w_{ij} = \eta ( <v_i h^*_j>_{data} - <v^*_i h^*_j>_{model})$
   \STATE set $w_{ij} = w_{ij} + \Delta w_{ij} $
   \ENDFOR
\end{algorithmic}
\end{algorithm}

To train a deep autoencoder using a stack of RBM, we sequentially train RBM that treat the previously generated hidden layer as the visible layer for the next RBM. More formally, if we define the conditional probabilities $p(v_i = 1 | h)$ and $ p(h_j = 1 | v) $ for the visible and hidden units of each RBM layer, we can draw samples from the conditional distribution in order to approximate $p(h)$ and $p(v)$ in order to train the RBM at each layer. After training one layer of our RBM stack, we set $p(v^{k+1}) = p(h^{k})$ for the hidden units from the previous layer, and train a new set of hidden units with the conditional probabilities  $p(h^{k+1} | h^{k})$. \\

After training the RBM stack, we use the weights generated by this pre-training approach as inputs to the back propagation algorithm as a fine-tuning step to further optimize the autoencoder structure \cite{hintonrbmpractical}. We can define our non-linear activation functions as sigmoidal functions so that probabilities are approximately maintained by setting $ g(a_j) = \sigma (a_j)$. The end result is an efficiently trained autoencoder that can be trained with somewhat less compute power than solely using backpropagation. \\

\subsection{Training autoencoders in practice}

The RBM stack approach to training a deep autoencoder is greedy. It can be shown that this approach does not exactly approximate the marginal log-likelihood. However, there are some theoretical variational bounds for the distribution learned by the stack of RBM compared to the true posterior that allow one to be more comfortable. The approach seems to work well in practice \cite{hintonrbmpractical}. However, it should be noted that \cite{semantichashing} is an older paper, and contemporary autoencoder training methodologies seem to have moved away from these older approaches. RBMs have seen less support in popular packages such as Tensorflow, Torch, and Keras. The more contemporary approach to training an autoencoder involves backpropagation variants with more effective activation functions such as ReLu, and faster optimization routines such as adam or adadelta. This approach is currently well supported by readily available packages, so it is more in line with existing software and hardware.  \\

\section{Semantic Hashing with autoencoders}

Semantic Hashing works through structuring an autoencoder such that the encoding layers of the network train an n-bit representation of our data at the narrowest layer of our network. In fact, the activations at this narrowest layer will be probabilities. But, we can translate these probabilities to binary representations. The n-bit encodings generated with the binary representations can be used as memory addresses so that we can easily retrieve all documents encoded within a hamming ball of arbitrary distance.  This is feasible do to the speed of known bit counting operations, which can be used to compute hamming distances efficiently. In this way, we reduce the number of cosine similarity computations needed for a query from the entire document set, to the preselection set of documents similarly indexed by the codes in our hamming ball. The documents within a hamming ball of distance $k$ are treated as a pre-filtered set of documents that are likely to be relevant. In practice, I find that the likelihood of similarly encoded documents  in the preselection set being relevant is generally higher than the corpus overall. This turns out to be a very useful property.  After the pre-filtering phase, the system can be treated as a standard information retrieval system over the documents in the preselection. Standard techniques such as TF-IDF can be used, but I also present some novel approaches in section 4. \\

TF-IDF with the semantic hashing preselection set outperforms vanilla TF-IDF.  We can think of the entire dataset as a distribution over the varying document categories with about equal probability. In the twenty news corpus of $n$ documents this turns out to be around a 5\% for each class. The preselection process generates a selection of documents that on average have a density that favours that class associated with the encoding document. In practice, I observed encoding densities of between 7\% to 12 \% for the $s$ preselected documents. If we consider the TF-IDF selection phase as an observation where selecting a document of the correct class is treated as a probability. Then the conditional probability for this selection, given that the document comes from the preselection set, will on average be higher than that of the full corpus. When we average the precision over all section queries, we essentially take the expectation of this conditional probability, which will be larger for the semantic hashing system. \\ 

The tuning of such a system is actually a fairly complicated engineering task. The performance of such a system is sensitive to several key tuning aspects that we will go over in this section.  \\ 

\subsection{Feature Selection}

The number of features chosen as input to an information retrieval system will affect both the speed of training of the autoencoder and the speed of the subsequent tfidf similarity computations. Increasing the number of features increases the number of parameters in the input and output layer of the neural network. This in turn increases the number of computations required to perform the gradient optimization routines used to train the network. Increasing the length of the feature vector increasing the burden of each cosine similarity computation required when ranking the top retrieved results in the TF-IDF phase. These computations represent one of the more demanding aspects of a traditional information retrieval system. Therefore, reducing the length of the feature vector is beneficial for the time performance of the system. \\

The trade off with reducing the number of features chosen is information loss, which can negatively impact the search results. Similarly the quality of features has a massive impact on the precision of the system. In \cite{semantichashing} the authors state that after trimming stopwords and stemming, the 20 most frequent words were used as features for their system. I think this is a simplification, particularly for the 20-newsgroups dataset. The quality of text in the 20-newsgroups dataset is fairly poor. More sophisticated techniques are required to remove distracting artifacts in the data such as names, email addresses, and inline symbols. Choosing a useful set of features is particularly important for the TF-IDF phase, as it is essentially useless if not harmful if bad features are chosen. \\

\subsection{Encoding Density}

The thinnest layer of the autoencoder is responsible for encoding the binary representations of each document. The width of this layer becomes an important tuning parameter, as it effectively controls the number of documents stored at each encoding address. A 20-bit encoding may pre-select thousands of similar documents while a 256-bit encoding may only find several. The width of the encoding layer of the network should be tuned to the size of the input set, similar to how a hash map should be tuned to prevent collisions. \\

The width of the encoding layer is directly linked to the precision and recall performance of the information retrieval system. If the width is large, the retrieval will be poor because only a few documents are preselected. In this case, it is even difficult to get an accurate evaluation of recall as simply too few documents are selected.  If the width of the thinnest layer is too small, then there are many collisions at each hash address. This reduces the speed performance gain as the number of cosine similarity computations will be high, and reduces the quality performance gain as the preselection set will not be meaningful. \\

Within a fixed width, the encodings created by the thinnest layer of the autoencoder become increasingly meaningful the more distinct they are. Enforcing this comes in the form of making sure that the activations at the encoding layer are as binary as possible. In other words, the activations for most vector should be sufficiently far from $ 0.5 $. Furthermore, if this is not the case, then rounding errors can lead to stochastic behaviour in the output encodings associated with each vector. We want to be able to control the density of relevant documents in the preselection set, and this stochastic behaviour leads to unpredictability. It is better to enforce more distinct activation probabilities at the encoding layer, and manage the density through the width of the encoding layer. In \cite{semantichashing} it is recommended to add gaussian noise during training in order to force the network to make stronger inferences. In practice, that means we should tune the gaussian noise added to the denoising autoencoder in order to enforce a consistent preselection density. \\

\subsection{Hamming Ball Radius}

The preselection phase of the semantic hashing system will iterate over the query documents to find those with an encoding within the hamming ball of radius $h$ around the encoding of the query. By increasing the hamming radius we increase the number of documents preselected. If our assumptions are correct that a document with an encoding similar to a query will be a similar document, then increasing the number of documents preselected via increase the hamming ball radius will not necessarily increase the number of \textit{relevant} documents. Subsequent TF-IDF ranking should sort out the irrelevant documents, but there is inherently a trade-off between with the quantity of documents preselected and the risk of irrelevant documents being selected by TF-IDF. The hamming radius should therefore be tuned in order to balance precision and recall. \\

Collecting documents with codes in the hamming ball of radius h can be approached in two ways. Either a generative approach can be taken, where the encodings within radius $h$ are generated and collected into a map, or the documents can be iterated over and a xor and bit count step can check to see whether a document's encodings is within the range of the queries'. The generative approach scales with the number of documents, but becomes intractable as h gets large. The selective approach scales with the hamming radius $h$, but $O(n)$ count and xor operations where $n$ is the number of documents.

\section{Information Retrieval} 

The core issue that the field of Information Retrieval aims to tackle is to extract the important information from a more voluminous collection of data. We can illustrate this problem with finding the most informative documents in a corpus, or, in our case, with finding similar documents or similar images to a given search query. The kind of systems that have been developed to solve these kinds of problems underly the modern information economy. These kinds of systems are generally evaluated on the quality of retrievals, but the speed is also an important factor. \\

\subsection{TF-IDF}

The most standard algorithm used to solve the information retrieval  tasks is the comparison of TF-IDF vectors. TF-IDF is an abbreviation of Term Frequency - Inverse Document Frequency. A TF-IDF transformation of a word count vector is a vector of TF-IDF scores computed for each word in the original vector. The transformation for each word $w$ with a count $c(w)$ in a word vector of length $W$ in a document collection $D$ is as follows, \\

\begin{equation}
TF_{w} = [tf(w)] * [idf(w)] = [\frac{c(w)}{W} ] * [-\log(\frac{ | D | }{ | D_w |} ]
\label{eq:gsaquery}
\end{equation}

where $D_w$ is the set of documents that contain w

\subsection{Pseudo-Relevance-Feedback}

A strategy for improving the performance of an information retrieval system is to transform queries using pseudo relevance feedback \cite{rocchiorelevance}. The initial query can be updated based on the results from the top k results in a first retrieval pass. For selection set of size k, we update the original query with 

\begin{equation}
q_{prf} = q + \frac{1}{k} \sum_{i=1}^k document_i
\label{eq:gsaquery}
\end{equation}

\subsection{Gradient Search Augmentation}

Here, I introduce a new method of using autoencoders to improve search accuracy. Recall the asymptotic behaviour $ \frac{r_\sigma(x)-x}{\sigma^2} \rightarrow \frac{\partial \log p(x)}{\partial x}$ as $\sigma \rightarrow 0$. If we have a trained autoencoder, we can update our search query $q$ 
with 

\begin{equation}
q_{gsa} = q-\frac{r(q)-q}{\sigma^2}
\label{eq:gsaquery}
\end{equation}

This object approximates $q - \nabla \log p(q)$. The geometric behaviour envisioned is that we move the original search query along the gradient in order to generate a query that is more probable given the manifold structure for $p(x)$ that we learned with our autoencoder. The goal is to make the information in the query more distinct, and therefore more likely to relate to documents in a similar class. This theory is validated in practice as we see that augmenting the query by the negative gradient generally outperforms augmenting the query by the positive gradient. 

Consider an alternative to GSA, a reconstruction search, where we compute the reconstruction $r(q)$ trained by our autoencoder, and then compute TF-IDF using our reconstruction rather than the original query. This could seem to work intuitively, but it actually performs very poorly. Even though the reconstruction $r(x)$ should be a more probable data point than the original data point x, the retrieved results generally do not agree with the query. What could be happening is that $r(x)$ becomes more probable at the cost of a loss of information that would make the search query unique. $r(x)$ is more probable relative to the \textit{entire} dataset, rather than the subset of documents in the proper category. By this reasoning, GSA can be seen as an analog to pseudo relevance feedback. Whereas pseudo relevance feedback introduces category bias from the top selected documents, GSA seems to remove bias of the query towards the averaged distribution learned by the autoencoder. \\

\section{Experiments}

In this section, I develop a formulation for how the performance of information retrieval systems can be measured. I discuss the results from several experiments with reconstruction search and gradient search augmentation (GSA), and try to motivate the reasoning for why GSA is successful. I evaluate both PRF and GSA augmentations in order to compare their respective benefits and differences. \\

\subsection{Evaluating information retrieval systems}

The standard metrics for evaluating the quality of search results are precision and recall. In the context of document retrieval, precision allows one to measure how frequently one selects relevant documents to a given query. Recall on the other hand allows one to measure how many of the relevant documents (true positives TP) one is capturing. It is usually the case that systems will attempt to balance precision and recall scores. Extending ones net to cover all documents can lead to false positives (FP), and tightening the constraints of the system to reduce the number of false positives, can lead to false negatives (FN). \\

 \begin{equation}
precision = \frac{TP}{TP+FP}
\label{eq:gsaquery}
\end{equation}

\begin{equation}
recall = \frac{TP}{TP+FN}
\label{eq:gsaquery}
\end{equation}

Because the number of documents retrieved in the preselection phase is variable at both a document and experiment level (based on a given system configuration). I find that the standard precision recall measurements are somewhat non-reflective of the actual system. Instead of precision and recall evaluations directly, I will compare the various search methodologies with the precision at each returned document count. For example precision at 5 for the precision in the first 5 documents retrieved, and precision at 10 for the first 10 documents. This approximates the characteristics of the precision recall tradeoff, but is more in line with the functionality of the semantic hashing system. \\

Usually manual annotations are used to measure whether two documents are similar. For our purposes we will use the class labels of each document to determine whether they are similar. So, if the document and the query have the same class label, then the document is considered relevant. \\

\subsection{Experiment Setup}

The twenty newsgroups dataset was used for this experiment. Text preprocessing on the dataset was used to alter and condense the number of features extracted from the corpus. The data is transformed into a vector of word counts for the various documents. Only words consisting of purely alphabetic characters that occur in less than 90\% and more than 0.001 \% of documents  are retained. English stop words such as 'the' and 'is' are also pruned. The 10000 most frequent words in the resulting corpus are retained, and this 10000 dimensional set of features is transformed into a TF-IDF vector. This transformation is applied to both the training and test dataset. \\ 

The network of the auto encoder follows a $[10000, 500, 500, 20, 500, 500, 10000]$ dimensional structure where the encoding and output layers are trained logistic units and the other layers are trained with ReLu activations. The autoencoder is trained as a denoising autoencoder where the training dataset is augmented with gaussian noise with $\mu=0$ and $\sigma = 2$. This approach was found to be simpler than pretraining with a stack of RBM, and more efficient as well as less prone to overfitting than training as a stack of purely logistic activations. The autoencoder is trained using Keras using the adadelta optimizer and the binary cross entropy loss function for 20 epoch with batches of size 256. \\ 

The several configurations of the search function for the Information Retrieval system are ran and their precisions over the first 100 documents are plotted and compared. There are some useful insights that can be derived from the varying functionality of the different systems. \\

\subsection{Information Retrieval performance of GSA and PRF}
\begin{figure}[ht]
\centering
\includegraphics[scale=0.5]{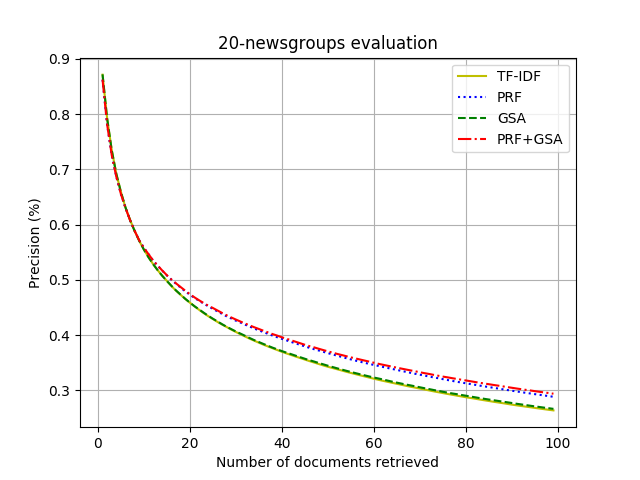}
\caption{Precisions of semantic hashing variants for the twenty news test dataset}
\label{fig:precision_results}
\end{figure}

GSA is found to increase the precision marginally over TF-IDF for both the early and later document retrievals. This improvement is not strongly pronounced, and depending on the adjustment can be as little as a 1\% increase. However, I do find across experiments that the performance increase is consistent across evaluations. \\

\begin{figure}[ht]
\centering
\includegraphics[scale=0.5]{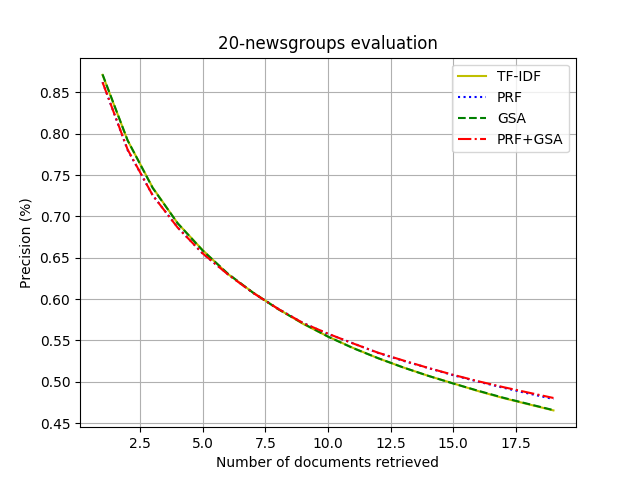}
\caption{Precisions of semantic hashing variants on the top retrievals of the twenty news test dataset}
\label{fig:precision_results}
\end{figure}

If we look at figure 3, we can see that the benefit of Pseudo Relevance feedback is seen in the later stages of retrieval. In figure 2, we see the top document precisions usually drops slightly compared to before these vectors are included, but we see average precision increase marginally over the later precisions. We can interpret this as a result of the information gain from including the top selected queries. There is a time tradeoff for this step since a retrieval of the first k documents requires iteration over the corpus, but the increase in quality of results can be worth it depending on the application. \\ 

\begin{figure}[ht]
\centering
\includegraphics[scale=0.5]{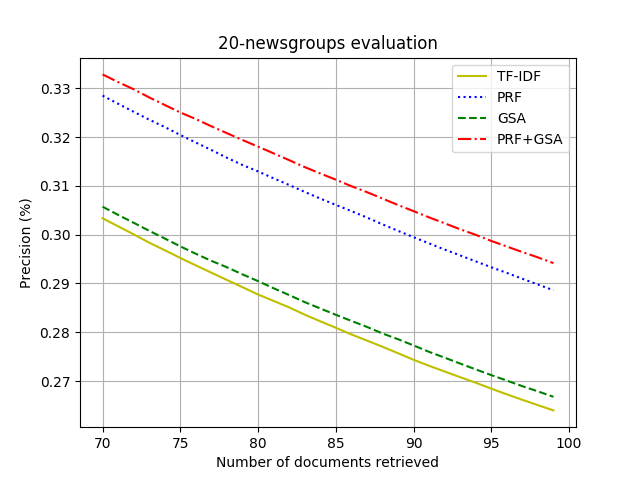}
\caption{Precisions of semantic hashing variants on the tail retrievals of the twenty news test dataset}
\label{fig:precision_results}
\end{figure}

GSA as well as PRF are found to improve performance, but there are tradeoffs to both approaches. GSA is more efficient than PRF since it does not require a second retrieval, whereas PRF does. On the other hand, PRF is seen to to produce a larger improvement in precision. \\

The combination of GSA+PRF performs best overall across precision values. It effectively combines the performance improvement from in early retrievals through GSA and carries this through later retrievals with PRF. The improvement in precision over the TF-IDF baseline is about 2-3. It is an interesting result that these two techniques are complimentary. \\

\section{Conclusion}

In this paper, we looked at the theory behind autoencoders in order to develop a justification for why they may help with information retrieval. We then considered the Semantic Hashing architecture, and outlined several key parameters that need engineering for the system to work to its full potential. We then observed the properties of several information retrieval search methods, including a novel search strategy, using a tuned semantic hashing system. I find that gradient search augmentation (GSA) and pseudo-relevance-feedback (PRF) both work sufficiently well in boosting performance over the standard TF-IDF strategy and the combination of both is best overall. \\

The caveat of this paper is that the performance of my trained system does not match the performance of the original system in \cite{semantichashing}. This is most likely the case due to the performance of the encoder function of the trained autoencoder. Additional tuning is probably required, or it may be the case that pre-training an autoencoder with a stack of RBM is more likely to generate a more desirable network structure for this task. \\

There are a few ideas that could be evaluated in future work. Unfortunately, there was not enough time to include these evaluations for this draft. Firstly, I would like to explore a form of PRF that uses data the autoencoder exclusively. To be explicit, we could augment the query with the $k$ documents retrieved in a hamming ball of distance 1 from the original encoding. Another area of exploration is to paramaterize the gradient update in GSA with $q_{gsa} = q-\alpha ( \frac{r(q)-q}{\sigma^2})$. Initial explorations of this possibility have shown that varying and cross-validating $\alpha$ can indeed help with performance. \\

Despite the performance gap found between this paper and \cite{semantichashing}, we can still see how the novel ideas contained in this paper contribute to an improved system relative to the baseline. The lessons learned from GSA are also relevant to note. We gain some insight into how the manifold structure of $p(x)$ can help us to improve an information retrieval system. A more intuitive understanding of why semantic hashing works for information retrieval was also presented. I find that the semantic hashing system is useful for information retrieval, but it should be noted that a lot of dedicated feature engineering and network tuning is required to get the system working effectively. Code is available at \cite{github2018}. \\

\bibliography{Autoencoders_and_Semantic_Hashing}
\bibliographystyle{icml2017}

\end{document}